\documentclass{IEEEtran4PSCC}

\usepackage{cite}

\ifCLASSINFOpdf
   \usepackage[pdftex]{graphicx}
\else
   \usepackage[dvips]{graphicx}
\fi

\usepackage[cmex10]{amsmath}
\usepackage{amssymb,mathtools}

\ifCLASSOPTIONcompsoc
 \usepackage[caption=false,font=normalsize,labelfont=sf,textfont=sf,position=top]{subfig}
\else
 \usepackage[caption=false,font=footnotesize,position=top]{subfig}
\fi

\usepackage{url}

\graphicspath{{./figures/}}
\usepackage{pgf}
\usepackage{siunitx}
\usepackage{tabularx}
\usepackage{tikz}
\usetikzlibrary{shapes,positioning,arrows,trees,fit}

\makeatletter
\let\old@ps@headings\ps@headings
\let\old@ps@IEEEtitlepagestyle\ps@IEEEtitlepagestyle

\usepackage{enumitem}                   %

\usepackage{booktabs}
\usepackage{eurosym}

\usepackage{cleveref}
\usepackage{placeins}
\crefname{equation}{}{}
\Crefname{equation}{Equation}{Equations}
\crefname{figure}{Fig.}{Figs.}
\Crefname{figure}{Fig.}{Figs.}
\crefname{table}{Table}{Tables}
\Crefname{table}{Table}{Tables}
\crefname{section}{section}{sections}
\Crefname{section}{Section}{Sections}
\crefname{algorithm}{Algorithm}{Algorithms}
\Crefname{algorithm}{Algorithm}{Algorithms}

\usepackage{multirow}
\usepackage[acronym]{glossaries}
\newacronym{der}{DER}{distributed energy resources}
\newacronym{ees}{EES}{electric energy storage}
\newacronym{mg}{MG}{microgrid}
\newacronym{mssp}{MSSP}{multi-stage stochastic programming}
\newacronym{vres}{VRES}{variable renewable energy source}
\newacronym{soc}{SOC}{state-of-charge}
\newacronym{dod}{DOD}{depth-of-discharge}
\newacronym{sddp}{SDDP}{stochastic dual dynamic programming}
\newacronym{pv}{PV}{photovoltaic}

\begin{document}
\title{Stochastic Operation of Energy Constrained Microgrids Considering Battery Degradation}

\author{
    \IEEEauthorblockN{
        Per Aaslid\IEEEauthorrefmark{1}\IEEEauthorrefmark{2},
        Magnus Korpås\IEEEauthorrefmark{2},
        Michael M Belsnes\IEEEauthorrefmark{1} and
        Olav B Fosso\IEEEauthorrefmark{2}
    }
    \IEEEauthorblockA{
        \IEEEauthorrefmark{1}Department of Energy Systems,
        SINTEF Energy Research, Trondheim, Norway
    }
    \IEEEauthorblockA{
        \IEEEauthorrefmark{2}Department of Electrical Power Engineering,
        Norwegian University of Science and Technology, Trondheim, Norway
    }
}

\maketitle

\begin{abstract}
Power systems with high penetration of variable renewable generation are vulnerable to periods with low generation. An alternative to retain high dispatchable generation capacity is electric energy storage that enables utilization of surplus power, where the electric energy storage contributes to the security of supply. Such systems can be considered as energy-constrained, and the operation of the electric energy storage must balance operating cost minimization and the risk of scarcity. In this matter, operation dependent storage characteristics such as energy storage degradation are a complicating factor. This paper proposes a linear approximation of battery state-of-charge degradation and implements it in a stochastic dual dynamic programming based energy-management model in combination with cycling degradation. The implications of degradation modelling are studied for a small Norwegian microgrid with variable renewable power generation and limited dispatchable generation capacity as well as battery and hydrogen storage to balance supply and demand. Our results show that the proposed strategy can prolong the expected battery lifetime by more than four years compared to the naive stochastic strategy but may cause increased degradation for other system resources. It is also evident that a stochastic strategy is crucial to retain low risk of scarcity in energy-constrained systems.
\end{abstract}


\thanksto{\noindent This work has been funded by the Norwegian Research Council under grant number 272398.}

\section{Introduction}
This paper addresses important operational challenges of future electricity systems with high \gls{vres} penetration and \gls{ees}, where new types of technological constraints will influence the optimal scheduling decisions. New methodologies that capture risk of scarcity, consequence of uncertainty, flexible demand, energy storage, and degradation are needed. The presented work investigates how degradation impacts operational costs, strategy, and the expected lifetime of batteries. These are crucial mechanisms and should be included in both the planning and operation of the future power system, where the installed solar \gls{pv} and wind capacity must be quadrupled by 2030 to follow the net zero pathway \cite{InternationalEnergyAgency2021a}.

Systems with high penetration of \glspl{vres} rely on sufficient dispatchable generation capacity to meet the peak demand and also in the hours with low \gls{vres} generation. An alternative to dispatchable thermal generation capacity is to utilize \gls{ees} flexibility. A challenge with \gls{ees} is that the current decision also affects the future energy content and the capability of providing capacity in the future. The decisions here and now must be taken while accounting for future power generation and load under uncertainty, and needs to balance the risk of generation curtailment versus the risk of scarcity. The operation of \gls{ees} in these situations can be seen as a precaution or arbitrage against extreme prices \cite{Geske2020,Aaslid2021}.

Different \gls{ees} technologies have complementary properties with respect to power and energy scalability. Lithium-ion batteries have gained high attention both in the research community as well as for power system applications due to their ability to deliver and absorb very high power almost instantaneously with low losses. They also have a relatively high energy to weight ratio compared to similar battery technologies. However, they are expensive to scale up with respect to energy compared to hydrogen, which can be stored in large tanks and scaled up at a relatively low cost. However, the cost of fuel cells and electrolyzers are still very high, and the round-trip efficiency is poor compared to batteries \cite{Pellow2015HydrogenAnalysis}.

Degradation characteristics also differ for batteries and hydrogen systems. The aging of hydrogen fuel cells are largely affected by start, stop and rapid ramping. However, the degradation has often been studied for vehicles that exhibit several cycles each hour \cite{Ahmadi2020TheVehicles}, while a grid connected fuel cell will operate with less frequent cycling. Moreover, degradation of fuel cells can also be related to dry membranes caused by limited operation, and modest operation can extend the expected lifetime compared to low operation \cite{Bidoggia2013EstimationCharacterization}. The degradation cost of the hydrogen system has therefore been neglected.

The degradation of lithium-ion batteries is closely related to operating conditions like charge/discharge power, \gls{dod}, \gls{soc}, temperature, and ampere throughput \cite{Wang2017a}. Energy management of \gls{vres} typically involves daily cycles, and the battery will rarely operate close to its maximum power capabilities. Moreover, the battery temperature will be controlled to ensure optimal operating conditions and minimal degradation. This paper will therefore consider degradation caused by \gls{dod} and \gls{soc}.

Experimental results show that the degradation rate of lithium-ion batteries increases with increasing \gls{dod}. Moreover, the degradation rate is also higher for high \gls{soc} \cite{Keil2016CalendarBatteries,Liu2020AnBatteries}, but very low \gls{soc} will also cause high degradation \cite{Vetter2005AgeingBatteries,Gao2018AgingCathode,Zhu2021InvestigationRanges,Laresgoiti2015}. The aging is therefore influenced by the operational pattern, and the optimal power dispatch largely depends on the battery's aging model \cite{Wang2020ImpactMicrogrid}. Previous studies of \gls{mg} economic dispatch often neglect the cost associated with degradation \cite{Shuai2019,Gil-Gonzalez2019}. Single factor models, considering degradation either as a function of power, \gls{soc}, \gls{dod}, or ampere hour throughput, are also widely adopted \cite{Wang2020ImpactMicrogrid}. For example, references \cite{Elkazaz2020,Nguyen2016StochasticCost,Fioriti2021Multi-yearMethodologies,Su2014} assume the degradation is proportional with energy throughput. More sophisticated models capture non-linear effects, either as a single factor model \cite{Ju2018ACosts} or combined models \cite{Koller2013DefiningSystem,Xu2018}.

However, non-linear models are often computationally intensive, especially for large-scale systems and stochastic problems. Convex and linear problem formulations reduce the computational burden, and enable utilization of decomposition techniques such as dual decomposition. Reference \cite{Xu2018a} proposes a piece-wise linear relaxation of the the non-linear \gls{dod} degradation, and shows that a cost reduction can be achieved by considering the cyclic degradation in the market clearing of a battery. However, linear approximation of the non-linear \gls{soc} degradation has gained less attention.

This paper proposes a piece-wise linear approximation of the battery \gls{soc} degradation effect, and demonstrates it in combination with linear \gls{dod} degradation \cite{Xu2018a} on a real \gls{mg} from the EU project REMOTE \cite{2021RemoteProject,Marocco2018}. The system is operating islanded, and is energy constrained since the backup generator is too small to cover the peak demand and the \gls{ees} must be operated to prevent load shedding in extreme situations. The optimal operation is considered using rolling horizon \cite{Maciejowski2002PredictiveConstraints} and \gls{sddp} \cite{Pereira1991}. The system is simulated for a whole year with rolling horizon using real observations from the \gls{mg} and scenarios generated from historical weather forecasts. Infinite horizon is embedded using cyclic Markov chains \cite{Dowson2020}, and the implication of including battery degradation will be studied with respect to the costs, \gls{vres} and \gls{ees} utilization as well as expected battery lifetime.

The remainder of the paper is organized as follows: \cref{sec:method} describes the rolling horizon simulation method as well as the \gls{sddp} algorithm, and derives the linear power system model including the \gls{soc} degradation model; \cref{sec:implementation} presents how the proposed method is implemented and the numerical values of the cases; \cref{sec:results} shows and discusses the results; and \cref{sec:conclusions} draws the conclusions and suggests future work.

\section{Method} \label{sec:method}
This section presents the rolling horizon stochastic energy-management model where the goal is to find the optimal control decisions $u_s$ for each stage $s$ for the in-going state $x_s$ and the uncertainty $\omega_s$.

\subsection{Rolling horizon simulation}
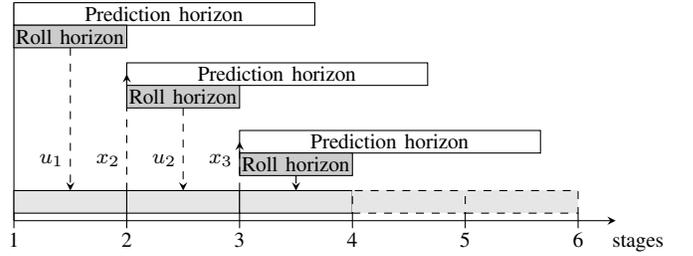
\begin{figure}[tb]
    \centering
    \begin{tikzpicture}
    \draw[-stealth] (0,0) -- (8,0);
    \draw[] (0,0) -- (0,2.9);
    \foreach \x in {1, ..., 6} {
        \draw[] (1.5*\x - 1.5, 0.1) -- (1.5*\x - 1.5, -0.1);
        \node[] at (1.5*\x-1.5, -0.25 ) {\footnotesize \x};
    }
    
    \draw[fill=white] (0,2.6) rectangle (4,2.9) node[midway]{\footnotesize Prediction horizon};
    \draw[fill=gray!40] (0,2.3) rectangle (1.5,2.6) node[midway]{\footnotesize Roll horizon};
    
    \draw[fill=white] (1.5,1.8) rectangle (5.5,2.1) node[midway]{\footnotesize Prediction horizon};
    \draw[fill=gray!40] (1.5,1.5) rectangle (3,1.8) node[midway]{\footnotesize Roll horizon};
    
    \draw[fill=white] (3,0.9) rectangle (7,1.2) node[midway]{\footnotesize Prediction horizon};
    \draw[fill=gray!40] (3,0.6) rectangle (4.5,0.9) node[midway]{\footnotesize Roll horizon};
    
    \draw[fill=gray!20] (0,0.1) rectangle (1.5,0.4) node[midway]{};
    \draw[fill=gray!20] (1.5,0.1) rectangle (3,0.4) node[midway]{};
    \draw[fill=gray!20] (3,0.1) rectangle (4.5,0.4) node[midway]{};
    \draw[fill=gray!20,dashed] (4.5,0.1) rectangle (6,0.4) node[midway]{};
    \draw[fill=gray!20,dashed] (6,0.1) rectangle (7.5,0.4) node[midway]{};
    
    \draw[-stealth, dashed] (0.75,2.3) -- (0.75,0.4);
    \node[] at (0.5,0.8) {\footnotesize $u_1$};
    \draw[-stealth, dashed] (2.25,1.5) -- (2.25,0.4);
    \node[] at (2,0.8) {\footnotesize $u_2$};
    \draw[-stealth, dashed] (3.75,0.6) -- (3.75,0.4);
    
    \draw[-stealth, dashed] (1.5, 0.4) -- (1.5, 1.95);
    \node[] at (1.25,0.8) {\footnotesize $x_2$};
    \draw[-stealth, dashed] (3, 0.4) -- (3, 1.05);
    \node[] at (2.75,0.8) {\footnotesize $x_3$};
    
    \node[] at (8.3,-0.3) {\footnotesize stages};
\end{tikzpicture}
    \vspace{-5mm}
    \caption{Rolling horizon optimization.}
    \label{fig:model_overview}
\end{figure}
Generation forecasts are based on weather forecasts issued with fixed intervals, and the stochastic model is trained each time a new forecast is available for the prediction horizon as illustrated in \cref{fig:model_overview}. The actual generation and demand is observed and evaluated for the roll horizon interval using the trained stochastic model yielding the optimal control $u_s$ that is implemented. Finally, the optimization horizon is moved forward and the procedure is repeated using the end state $x_{s+1}$ from previous optimization as the initial value.

\subsection{Multi-stage stochastic programming}
This paper considers \gls{mssp} for solving the proposed energy management problem. \Gls{mssp} captures that energy management is a sequential decision making process and recognizes that decisions can be updated stage-wise as uncertainty is revealed. The \gls{mssp} formulation in \cref{eq:value_func,eq:state_transition,eq:admissible_controls} is divided into several stages $s$, each representing a discrete moment in time. The goal is to minimize the current and future operating costs. State variables $x_s$ represent variables connected across stages, such as \gls{ees} \gls{soc}. Control variables $u_s$ are decisions, both implicit and explicit, and must be within the technical limitations of the system given by the set of admissible controls \cref{eq:admissible_controls}. The state transition function \cref{eq:state_transition} describes the relation between the state variables across stages. The random variable $\omega_s$ represents the uncertainty in demand and \gls{vres} generation.
\begin{subequations}
    \begin{multline}
        \min_{u_t} \Bigg\{ C_1(x_1, u_1, \omega_1) 
        + \underset{\omega_{2} | \omega_1}{\mathbb{E}} \bigg[ \underset{u_2}{\min} \Big( C_2(x_2, u_2, \omega_2) + \dots \\
        + \underset{\omega_S | \omega_{S-1}, \dots, \omega_2}{\mathbb{E}} \Big[ \underset{u_S}{\min} \big( C_S(x_S, u_S, \omega_S) \big) \Big] \Big) \bigg] \Bigg\}
        \label{eq:value_func}
    \end{multline}
    \begin{align}
        \text{s.t.} \quad x_{s+1} &= T_s(x_s, u_s, \omega_s) \label{eq:state_transition}\\
        u_s &\in U_s(x_s, \omega_s) \label{eq:admissible_controls} %
    \end{align} 
\end{subequations}

By assuming the random variable is stage-wise independent and has a discrete set of realizations for each stage, the problem can be formulated on extended form. Instead of solving the intractable extended problem, \gls{sddp} \cite{Pereira1991} decomposes the problem into sub-problems for each stage $s \in S$ and realization of the random variable $\omega_s \in \Omega_s$ as shown in \cref{eq:value_func_k,eq:dummy_k,eq:state_transition_k,eq:admissible_k,eq:cut_k}. The algorithm is divided into two phases: forward pass and backward recursion. The forward pass samples a random variable for each stage and solves the sequence of sub-problems using the outgoing state of stage $s$ as the in-going state to stage $s+1$. When the final stage is reached, the backward recursion starts by solving the final stage for all the random variables. From convexity, the dual variables $\lambda_s$ of the state variable $x_s$ \cref{eq:dummy_k} can be used to generate a linear cutting plane \cref{eq:cut_k} that acts as a lower bound for the previous stage problem, and the procedure is repeated all the way back to the first stage. The whole procedure is repeated, adding new cuts for each iteration $k$, until the convergence criteria is met \cite{Pereira1991,Dowson2020}.
\begin{subequations}
    \begin{align}
        \underset{u_s, x_s, x_{s+1}, \theta_s}{\min} &C_s(x_s, u_s, \omega_s) + \theta_s \label{eq:value_func_k}\\
        \text{s.t. } x_s &= \bar{x_s}, \quad [\lambda_s] \label{eq:dummy_k}\\
        x_{s+1} &= T_s(x_s, u_s, \omega_s) \label{eq:state_transition_k}\\
        u_s &\in U_s(x_s, \omega_s) \label{eq:admissible_k}\\
        \theta_s &\geq \alpha_s^k + \beta_s^k x_{s+1}, \quad k \in \{ 1,2, \dots , K \} \label{eq:cut_k}
    \end{align}
\end{subequations}

\subsection{Power system model} \label{sec:power_system_model}
The stage-wise power system model including the battery degradation model will be expressed using the following sets and indices:
\begin{itemize}
    \item $t \in \mathcal{T}$: Time index $t$ and the set of time steps $\mathcal{T}$.
    \item $g \in \mathcal{G}$: Generator $g$ and the set of dispatchable generators $\mathcal{G}$.
    \item $r \in \mathcal{R}$: \Gls{vres}  $r$ and the set of \glspl{vres} $\mathcal{R}$.
    \item $d \in \mathcal{D}$: Consumer $d$ and the set of consumers $\mathcal{D}$.
    \item $e \in \mathcal{E}$: \Gls{ees} $e$ and the set of \glspl{ees} $\mathcal{E}$.
    \item $k_\delta \in \mathcal{K}_\delta$: Battery cycling segment $k_\delta$ in the set of segments $\mathcal{K}_\delta$.
    \item $k_\sigma^{up} \in \mathcal{K}_{\sigma}^{up} / k_\sigma^{dn} \in \mathcal{K}_{\sigma}^{up}$: Battery \gls{soc} segment $k_\sigma^{up}/ k_\sigma^{dn}$ direction up/down in the set of segments $\mathcal{K}_\sigma^{up} / \mathcal{K}_\sigma^{up}$ respectively.
\end{itemize}
The following parameters have been used:
\begin{itemize}
    \item $\Delta T_t$: Timestep $t$ step length.
    \item $PG_g^{max}$: Generator $g$ maximum power dispatch.
    \item $C_g$: Generator $g$ marginal operating cost.
    \item $PR_{r,t}^{max}$: \Gls{vres} $r$ maximum power at time $t$.
    \item $PD_{d,t}$: Demand of consumer $d$ at time $t$.
    \item $C_d$: Consumer $d$ marginal load shedding cost.
    \item $SOC_e^{min} / SOC_e^{max}$: \Gls{ees} $e$ minimum and maximum \gls{soc}.
    \item $PS_e^c / PS_e^d$: \Gls{ees} $e$ maximum charge/discharge power.
    \item $C_{e, k_\delta}$: \Gls{ees} $e$ \gls{dod} marginal degradation cost segment $k$.
    \item $SOC_e^{ref}$: \Gls{ees} $e$ \gls{soc} degradation reference value.
    \item $C_{e, k_\sigma}^{up} / C_{e, k_\sigma}^{dn}$: \Gls{ees} $e$ \gls{soc} marginal degradation cost up/down segment $k_\sigma^{up} / k_\sigma^{dn}$.
    \item $\eta_e^c / \eta_e^d$: \Gls{ees} $e$ charge/discharge efficiency.
    \item $R$: Battery replacement cost.
\end{itemize}
The functions and variables are:
\begin{itemize}
    \item $p_{g,t}$: Generator $g$ power dispatch at time $t$.
    \item $p_{g,t}$: \Gls{vres} $r$ power dispatch at time $t$.
    \item $p_{d,t}$: Consumer $d$ demand at time $t$.
    \item $pls_{d,t}$: Consumer $d$ load shedding at time $t$.
    \item $ps_{e,t(,k_\delta)}^c / ps_{e,t(,k_\delta)}^d$: \Gls{ees} $e$ power charge/discharge \gls{dod} (segment $k_\delta$) at time $t$.
    \item $soc_{e,t,k_\delta}$: \Gls{ees} $e$ \gls{soc} \gls{dod} segment $k_\delta$ at time $t$.
    \item $soc_{e,t,k_\sigma}^{up} / soc_{e,t,k_\sigma}^{dn}$: \Gls{ees} $e$ \gls{soc} degradation segment $k_\sigma^{up} / k_\sigma^{dn}$ above/below reference value at time $t$.
    \item $\delta_t$: Unitless \gls{ees} cycle depth at time $t$.
    \item $\sigma_t$: Unitless \gls{ees} \gls{soc} at time $t$.
    \item $\sigma^{ref}$: Unitless \gls{ees} reference \gls{soc}.
    \item $f_\delta(\delta_t)$: Incremental battery fade as a function of cycle depth $\delta_t$ at time $t$.
    \item $f_\sigma(\sigma_t)$: Incremental battery fade as a function of \gls{soc} $\sigma_t$ at time $t$.
\end{itemize}

The resulting model is summarized in \cref{eq:objective,eq:power_balance,eq:thermal_gen,eq:vres_gen,eq:demand,eq:charge_seg_limit,eq:discharge_seg_limit,eq:charge_sum,eq:discharge_sum,eq:energy_balance,eq:storage_segment_size}. The objective is to minimize the dispatchable generation costs, load shedding, and \gls{ees} degradation associated with both \gls{dod} and \gls{soc} \cref{eq:objective}. The total power injections and withdrawals must balance at all time steps \cref{eq:power_balance}. The generation, both dispatchable and \gls{vres}, must respect the maximum generation \cref{eq:thermal_gen,eq:vres_gen}. Demand that can not be met causes load shedding \cref{eq:demand}. The battery charge/discharge must respect the maximum limits, both per segment \cref{eq:charge_seg_limit,eq:discharge_seg_limit} and the sum of the segments \cref{eq:charge_sum,eq:discharge_sum}. The \gls{ees} energy balance is expressed per segment \cref{eq:energy_balance}, where the segments are divided into equal sizes \cref{eq:storage_segment_size}.
\begin{multline}
    \min \sum_{t \in \mathcal{T}} \Bigg[
        \sum_{g \in \mathcal{G}} C_g p_{g,t}
        + \sum_{d \in \mathcal{D}} C_d pls_{d,t}
        + \sum_{e \in \mathcal{E}} \sum_{j \in \mathcal{J}}C_{e, k_\delta} ps_{e,t,k_\delta}^d \\
        + \sum_{e \in \mathcal{E}}
        \bigg( \sum_{k_\sigma \in \mathcal{K}_\sigma^{up}} C_{e, k_\sigma}^{up} soc_{e,t,k_\sigma}^{up} + \sum_{k_\sigma \in \mathcal{K}_\sigma^{dn}} C_{e, k_\sigma}^{dn} soc_{e,t,k_\sigma}^{dn} \bigg)
    \Bigg] \label{eq:objective}
\end{multline}
\vspace*{-2mm}
$\text{subject to}$
\begin{multline}
    \sum_{g \in \mathcal{G}} p_{g,t}
    + \sum_{r \in \mathcal{R}} p_{r,t}
    + \sum_{e \in \mathcal{E}} ps_{e,t}^d %
    =
    \sum_{d \in \mathcal{D}} p_{d,t}
    + \sum_{e \in \mathcal{E}} ps_{e,t}^c
    \label{eq:power_balance}
\end{multline}
\vspace*{-5mm}
\begin{align}
    0 &\leq p_{g,t} \leq PG_g^{max} \label{eq:thermal_gen} \\
    0 &\leq p_{r,t} \leq PR_{r,t}^{max} \label{eq:vres_gen} \\
    p_{d,t} &= PD_{d,t} - pls_{d,t} \label{eq:demand} \\
    0 &\leq ps_{e,t,k_\delta}^c \leq PS_e^c\label{eq:charge_seg_limit}\\
    0 &\leq ps_{e,t,k_\delta}^d \leq PS_e^d \label{eq:discharge_seg_limit}\\
    ps_{e,t}^c &= \sum_{j \in \mathcal{J}} ps_{e,t,k_\delta}^c \leq PS_e^c \label{eq:charge_sum} \\
    ps_{e,t}^d &= \sum_{j \in \mathcal{J}} ps_{e,t,k_\delta}^d \leq PS_e^d \label{eq:discharge_sum}
\end{align}
\vspace*{-5mm}
\begin{multline}
    soc_{e,t,k_\delta} = soc_{e,t-1,k_\delta} \\ + \Delta T_t \left( \eta_e^c ps_{e,t,k_\delta}^c - \frac{1}{\eta_e^d} ps_{e,t,k_\delta}^d \right) \label{eq:energy_balance}
\end{multline}
\vspace*{-5mm}
\begin{align}
    0 &\leq soc_{e,t,k_\delta} \leq \frac{1}{|\mathcal{K}_\delta|} (SOC_e^{max} - SOC_e^{min}) \label{eq:storage_segment_size}
\end{align}

Note that restrictions to prevent simultaneous charging and discharging have not been included, which implies that dumping of energy from the battery is accepted. This is not a problem when the \gls{vres} generation can be curtailed at no cost. However, by introducing \gls{soc} degradation cost, situations where dumping of energy is beneficial might arise.

The state $x_s$ comprises the initial \gls{soc} variable $soc_{e,t,k_\delta}$ at each stage, and the final \gls{soc} variable at the outgoing state $x_{s+1}$. The random variable $\omega$ comprises the renewable generation $PR_{r,t}^{max}$ and the demand $PD_{d,t}$ for all the steps in the stage. The remaining variables are decisions $u_s$, either explicit or implicit.

\subsection{\gls{ees} degradation model} \label{sec:ees_degradation}
The proposed model assumes the degradation due to \gls{dod} and \gls{soc} are decoupled. For an arbitrary convex \gls{dod} capacity fade function $f_{\delta}(\delta_t)$, the \gls{ees} \gls{soc} is divided into equally sized segments $K_\delta$ yielding cost coefficients $C_{\delta k}$ \cref{eq:linear_cost_coeff} \cite{Xu2018a}.

\begin{align}
    C_{\delta k} = \frac{R}{\eta^d SOC^{max}} |\mathcal{K}_\delta| \left[ f_\delta \left( \frac{k}{|\mathcal{K}_\delta|} \right) - f_\delta \left( \frac{k-1}{|\mathcal{K}_\delta|} \right)  \right], k \in \mathcal{K}_\delta \label{eq:linear_cost_coeff}
\end{align}
Each energy level $soc_t$ and charge/discharge $ps_t^c / ps_t^d$ is divided into $K_\delta$ segments. Since the marginal cost curve is convex, the cheapest available segment will always be discharged, and the suggested method will therefore count cycles in a similar manner as the Rainflow counting algorithm \cite{Amzallag1994StandardizationAnalysis}.

For an arbitrary convex \gls{soc} capacity fade function $f_\sigma(\sigma)$, $\sigma^{ref}$ represents the \gls{soc} level where the \gls{soc} degradation is lowest as shown in \cref{eq:sigma_ref}.
\begin{align}
    \sigma^{ref} = \underset{\sigma}{\mathrm{argmin}} \frac{\partial f_\sigma(\sigma)}{\partial \sigma} \label{eq:sigma_ref}
\end{align}
The incremental capacity fade as a function of \gls{soc} can be found by taking the derivative of $f_{\sigma}(\sigma)$ with respect to $soc_t$.
\begin{align}
    \frac{\partial f_\sigma(\sigma_t)}{\partial soc_t} = \frac{d f_\sigma(\sigma_t)}{d \sigma_t} \frac{\partial \sigma_t}{\partial soc_t} = \frac{1}{SOC^{max}} \frac{d f_\sigma(\sigma_t)}{d \sigma_t}
\end{align}
The $soc_{e,t}$ variable is divided into $K_{\sigma}^{up}$ and $K_{\sigma}^{dn}$ equally sized segments $soc_{e,t,k}^{up}$ and $soc_{e,t,k}^{dn}$ for up and down direction, respectively, as shown in \cref{eq:soc_seg_up,eq:soc_seg_dn,eq:soc_seg_up_limits,eq:soc_seg_dn_limits} such that $soc_{e,t,k}^{up}$ and $soc_{e,t,k}^{dn}$ represent the distance from the reference value in both directions.
\begin{align}
    \sum_{k \in \mathcal{K}_\sigma^{up}} soc_{e,t,k}^{up} &\geq soc_{e,t} - SOC_e^{ref} \label{eq:soc_seg_up}\\
    \sum_{k \in \mathcal{K}_\sigma^{dn}} soc_{e,t,k}^{dn} &\geq SOC_e^{ref} - soc_{e,t} \label{eq:soc_seg_dn}\\
    0 \leq soc_{e,t,k}^{up} &\leq \frac{1}{|\mathcal{K}_\sigma^{up}|} \left( SOC^{max} - SOC_e^{ref} \right) \label{eq:soc_seg_up_limits}\\
    0 \leq soc_{e,t,k}^{dn} &\leq \frac{1}{|\mathcal{K}_\sigma^{dn}|} SOC_e^{ref} \label{eq:soc_seg_dn_limits}
\end{align}

Let $C_{\sigma,k}^{up}$ and $C_{\sigma,k}^{dn}$ denote the incremental aging cost with respect to each segment in either direction $soc_{e,t,k}^{up}$ and $soc_{e,t,k}^{dn}$. The resulting cost coefficients are expressed in \cref{eq:soc_linear_cost_coeff_up,eq:soc_linear_cost_coeff_dn}.

\begin{multline}
    C_{\sigma k}^{up} = \frac{R}{SOC^{max}} |\mathcal{K}_\sigma^{up}| \bigg[ f_\sigma \left( \sigma^{ref} + \frac{k}{|\mathcal{K}_\sigma^{up}|} (1-\sigma^{ref}) \right) \\
    - f_\sigma \left(\sigma^{ref} + \frac{k-1}{|\mathcal{K}_\sigma^{up}|} (1 - \sigma^{ref}) \right)  \bigg], k \in \mathcal{K}_\sigma^{up} \label{eq:soc_linear_cost_coeff_up}
\end{multline}
\vspace*{-5mm}
\begin{multline}
    C_{\sigma k}^{dn} = \frac{R}{SOC^{max}} |\mathcal{K}_\sigma^{dn}| \bigg[ f_\sigma \left( \sigma^{ref} - \frac{k}{|\mathcal{K}_\sigma^{dn}|} \sigma^{ref} \right) \\
    - f_\sigma \left(\sigma^{ref} - \frac{k-1}{|\mathcal{K}_\sigma^{dn}|} \sigma^{ref} \right)  \bigg], k \in \mathcal{K}_\sigma^{dn} \label{eq:soc_linear_cost_coeff_dn}
\end{multline}
The cheapest segments will always be used first, and the correct segment will be used given a convex cost function.

\section{Implementation and numerical values} \label{sec:implementation}
The proposed method is implemented in Julia (1.4.2) with SDDP.jl (0.3.14) \cite{Dowson2021} and Gurobi (9.1). The models were trained with 50 \gls{sddp} iterations.
\subsection{Microgrid}
\begin{figure}
    \centering
    \input{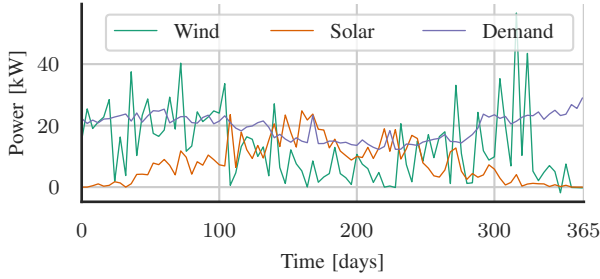}
    \caption{Four day average generation and demand for the optimization period.}
    \label{fig:gen_demand}
\end{figure}

The model has been tested on the Rye microgrid in central Norway that is partly funded by the Horizon 2020 project REMOTE \cite{2021RemoteProject}. The \gls{mg} comprises a few farms and houses, and is supplied by a wind turbine, solar \gls{pv}, and a diesel generator that serves as backup in case of insufficient \gls{vres} generation. The system is equipped with a battery and hydrogen storage to balance supply and demand \cite{Marocco2018}. Load shedding costs occur if the supply is unable to meet the demand. Numerical values for the \gls{mg} are shown in \cref{tab:mg_numerical_values}, and the \gls{ees} in \cref{tab:system_storages}. Note that the wind and diesel generator sizes in this case differs from the actual system. \Cref{fig:gen_demand} shows the four day average \gls{vres} generation and demand for the whole period.

\begin{table}[tb]
    \caption{Microgrid numerical values}
    \label{tab:mg_numerical_values}
    \centering
    \begin{tabular}{llr}
        \toprule
         Description & Unit & Value  \\
         \midrule
         Wind turbine capacity & [kW] & 135 \\
         Solar PV capacity & [kW] & 86 \\
         Diesel generator capcity & [kW] & 25 / 75 \\
         Diesel generation cost & [\euro/MWh] & 100 \\
         Load shedding cost & [\euro/MWh] & 5000 \\
         \bottomrule
    \end{tabular}
\end{table}
\begin{table}[tb]
    \caption{Numerical values for microgrid {ees}.}
    \label{tab:system_storages}
    \centering
    \begin{tabular}{llrr}
        \toprule
        Description & Unit & Lithium-ion & Hydrogen  \\
        \midrule
        Charge power & [kW] & 500 & 55\\
        Discharge power & [kW] & 500 & 100\\
        Size & [kWh] & 500 / 1000 & 3300 / -\\
        Charge efficiency & [\%] & 96 & 64 \\
        Discharge efficiency & [\%] & 96 & 50\\
        Replacement cost & [\euro/kWh] & 100 & NA \\
        \bottomrule
    \end{tabular}
\end{table}

\subsection{Battery degradation}
\begin{figure}[tb]
    \centering
    \input{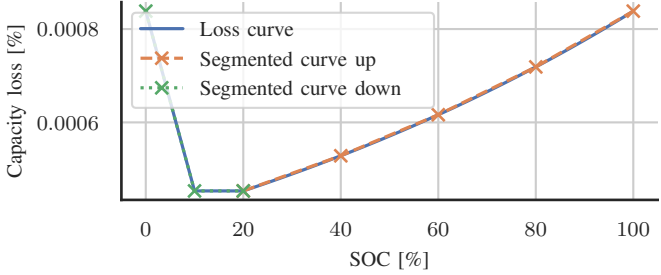}
    \caption{Continuous and linearized \gls{soc} capacity loss functions.}
    \label{fig:fade_functions}
\end{figure}

This paper uses a quadratic \gls{dod} capacity fade function \cref{eq:dod_stress} \cite{Laresgoiti2015,Koller2013DefiningSystem}. The upper range of the \gls{soc} stress function \cref{eq:soc_stress} is exponential \cite{Millner2010ModelingVehicles}, while the lower part is defined to capture the potential collapse associated with operating at very low \gls{soc} \cite{Xu2018,Zhu2021InvestigationRanges}. Note that any convex fade function can be used.
\begin{align}
    f_{\delta}(\delta) &= k_{\delta} \ \delta^2 \label{eq:dod_stress} \\
    f_{\sigma}(\sigma) &= 
    \begin{cases}
        k_{\sigma 1} e^{k_{\sigma 2} (\sigma - \sigma^{ref})} & 0.2 \leq \sigma \leq 1 \\
        f_\sigma(1.0) + \frac{\sigma}{0.1}(f_\sigma(0.2) - f_\sigma(1.0)) & 0 \leq \sigma < 0.1 \\
        f_\sigma(0.2) & 0.1 \leq \sigma < 0.2
    \end{cases}
    \label{eq:soc_stress}
\end{align}
Assuming the battery fade is 85\% higher at 90\% compared to 10\% \gls{soc} \cite{Stroe2015DegradationProfile}, yields $k_{\sigma 2} = 0.769$. The battery is assumed to reach end of life after 10 years with 3,000 cycles at 80\% \gls{dod} with 50\% average \gls{soc} and after 20 years with no cycling at 50\% \gls{soc}. These assumptions yield $k_\delta =\text{\num{3.092e-4}}$ and $k_{\sigma 1} = \text{\num{5.708e-6}}$, respectively. The resulting \gls{soc} loss function and the corresponding linearized segments are shown in \cref{fig:fade_functions}. The \gls{soc} loss curve is assumed to be flat between 10 and 20\%, and the \gls{soc} degradation is equal at 0 and 100\% \gls{soc} to also capture capacity fade at low \gls{soc}.

\subsection{Scenarios and stages}
The time-series for wind and solar \gls{pv} power and demand are based on actual values from the Rye \gls{mg} through year 2020 and is available online \cite{TrnderEnergi2021AIMicrogrid}. The wind power has been scaled down to 60\% of the original size. Wind, solar \gls{pv} and demand have each been forecasted with a low, medium, and high scenario with probability 20, 60 and 20\% respectively, based on the 0.2, 0.5, and 0.8 quantiles, yielding a combination of 27 scenarios. These have been ordered based on accumulated net production and reduced by selecting the median scenario of the percentiles 0-10, 10-30, 30-70, 70-90, and 90-100 with the corresponding probabilities 0.1, 0.2, 0.4, 0.2, and 0.1. More sophisticated scenario generation methods could have been used but are outside the scope of this paper. The individual percentiles are published online \cite{Aaslid2021RyeSet}. The four first stages are 6 hours each, the next is 24 hours, while the final is 72 hours and a repeated cyclic with discount factor 0.7 as described in reference \cite{Dowson2020}. The final stage is beyond the meterological forecast, and historical daily mean values are applied as scenarios using the same quantiles as previous stages.

\subsection{Cases}
The model has been simulated with three different variations of the system with respect to generator and \gls{ees} sizes as shown in \cref{tab:cases}. The systems in cases \textit{1} and \textit{3} are energy constrained since the diesel generator is too small to meet the peak demand, hence load shedding depends on how the \glspl{ees} are scheduled. However, the diesel generator in case \textit{2} is large enough to always meet the peak demand, hence there is never a risk of scarcity unless the generator fails.
\begin{table}[tb]
    \caption{Overview of cases the proposed model has been tested on.}
    \label{tab:cases}
    \centering
    \begin{tabular}{cccc}
        \toprule
        Case & Diesel generator [kW] & Battery [kWh] & Hydrogen [kWh] \\
         \midrule
         1. & 25 & 500 & 3300 \\
         2. & 75 & 500 & 3300 \\
         3. & 25 & 1000 & - \\
         \bottomrule
    \end{tabular}
\end{table}

Each case will be analyzed with perfect forecast, deterministic forecast, and stochastic optimization, both with and without degradation in the optimization model as shown in \cref{tab:methods}.
\begin{table}[tb]
    \caption{Overview of methods the proposed cases have been analyzed with.}
    \label{tab:methods}
    \centering
    \begin{tabular}{rlcc}
        \toprule
         Method & Forecast & DOD & SOC \\
         \midrule
         a. & Perfect & X & X\\
         b. & Deterministic &  & \\
         c. & Stochastic &  & \\
         d. & Stochastic & X & \\
         e. & Stochastic &  & X\\
         f. & Stochastic & X & X\\
         \bottomrule
    \end{tabular}
\end{table}

\section{Results and discussion} \label{sec:results}
The main objective is to always meet the demand in the most cost effective way by using as much \gls{vres} generation as possible, and only using diesel if necessary to avoid load shedding. \Gls{ees} must also be utilized to maximize the \gls{vres} utilization and to minimize the diesel consumption and the load shedding. However, the battery degradation is a complicating element. Although cycling the battery is less expensive than generating power from the diesel generator, even for deep cycles, it is difficult due to the uncertainty in determining if the power charged now is needed later or if it can be consumed directly from \gls{vres} generation. It is therefore necessary to balance the cost of cycling the battery toward the expected diesel generation reduction. Additionally, there is an increasing cost associated with staying at high \gls{soc}. It can therefore be cost effective to keep the \gls{soc} low in periods with a stable high \gls{vres} generation to extend the battery's lifetime.

\begin{table*}[tb]
\caption{Overview of operating costs, degradation costs, battery expected lifetime, VRES generation and EES charge/discharge for cases in \cref{tab:cases} and methods in \cref{tab:methods}.}
\label{tab:case_cost_summary}
\centering
\begin{tabular}{ccccccccccccc}
\toprule
&& \multicolumn{3}{c}{Operating cost $[\euro]$} & \multicolumn{3}{c}{Degradation cost $[\euro]$} & $[year]$ & \multicolumn{3}{c}{Energy $[MWh]$} \\
\cmidrule(lr){3-5} \cmidrule(lr){6-8} \cmidrule(lr){9-9} \cmidrule(lr){10-12}
Case & Method & Total & Load shedding & Diesel & DOD & SOC up & SOC down & Lifetime & VRES & H2 ch/dch & Batt ch/dch\\
\midrule
1 & a & 3719.7 & 0.00 & 3076 & 514 & 117 & 12 & 21.68 & 185.03 & 61.52 / 20.19 & 39.83 / 36.90\\
1 & b & 8541.1 & 4054.93 & 2840 & 1041 & 435 & 171 & 17.81 & 171.41 & 40.74 / 13.54 & 49.34 / 45.67\\
1 & c & 6011.1 & 968.94 & 2860 & 1328 & 731 & 123 & 16.26 & 166.73 & 32.87 / 11.02 & 52.67 / 48.73\\
1 & d & 5579.5 & 1086.00 & 2934 & 627 & 882 & 50 & 18.09 & 181.56 & 57.16 / 18.79 & 40.84 / 37.83\\
1 & e & 5804.1 & 1108.53 & 2861 & 1488 & 310 & 37 & 17.23 & 181.25 & 53.03 / 17.47 & 63.36 / 58.58\\
1 & f & 5256.8 & 1414.40 & 2964 & 582 & 268 & 29 & 20.62 & 183.48 & 60.64 / 19.90 & 39.71 / 36.79\\
\midrule
2 & a & 3719.7 & 0.00 & 3076 & 514 & 117 & 12 & 21.68 & 185.03 & 61.52 / 20.19 & 39.83 / 36.90\\
2 & b & 4508.7 & 0.00 & 2917 & 998 & 445 & 149 & 17.98 & 171.41 & 40.86 / 13.58 & 47.87 / 44.31\\
2 & c & 5093.3 & 0.00 & 2864 & 1285 & 723 & 220 & 16.13 & 166.46 & 32.50 / 10.90 & 50.46 / 46.69\\
2 & d & 4575.3 & 0.00 & 2892 & 650 & 848 & 185 & 17.69 & 181.75 & 56.55 / 18.60 & 40.43 / 37.45\\
2 & e & 4704.9 & 0.00 & 2883 & 1494 & 313 & 16 & 17.27 & 181.26 & 53.10 / 17.49 & 62.37 / 57.63\\
2 & f & 3814.1 & 0.00 & 3000 & 556 & 253 & 6 & 20.90 & 183.46 & 60.79 / 19.95 & 38.50 / 35.63\\
\midrule
3 & a & 4631.9 & 0.00 & 3459 & 771 & 348 & 53 & 19.45 & 141.21 & - / - & 56.93 / 52.66\\
3 & b & 9718.3 & 3822.67 & 3071 & 1228 & 1258 & 338 & 14.72 & 143.96 & - / - & 63.49 / 58.71\\
3 & c & 7202.8 & 775.38 & 3064 & 1436 & 1741 & 186 & 13.64 & 143.42 & - / - & 59.47 / 55.00\\
3 & d & 6968.8 & 1061.56 & 3091 & 885 & 1745 & 186 & 14.74 & 143.05 & - / - & 59.03 / 54.59\\
3 & e & 8467.5 & 1107.20 & 3215 & 3133 & 945 & 67 & 12.32 & 161.68 & - / - & 181.47 / 167.43\\
3 & f & 6188.1 & 1236.83 & 3321 & 776 & 789 & 66 & 17.86 & 140.56 & - / - & 56.43 / 52.20\\
\bottomrule
\end{tabular}
\end{table*}
The results in \cref{tab:case_cost_summary} show that accounting for \gls{dod} and \gls{soc} degradation increases the load shedding case \textit{1} and \textit{3}, and the diesel cost for all cases. However, the reduction in degradation surpasses the increase in diesel and load shedding costs and indicates more than four years of increase in expected lifetime for all cases when comparing methods \textit{c} and \textit{f}. A very common way to reduce battery degradation is to apply fixed operating limits, such as enforcing a permanent operating range between 10 and 90\%. However, in situations where the only alternative is load shedding, it is optimal to utilize the full battery range since the cost reduction associated with reducing load shedding outperforms the cost accrued by degradation.

The energy balance in cases \textit{1} and \textit{2} show that the hydrogen system replaces some of the battery cycling, since it has no degradation costs. However, case \textit{3}, which has no hydrogen in the system, also shows a significant cost reduction associated with battery degradation. However, \gls{soc} degradation without \gls{dod} degradation (method \textit{e}) causes significant cycling and energy dumping through simultaneous charging and discharging.

\begin{figure}[tb]
    \centering
    \input{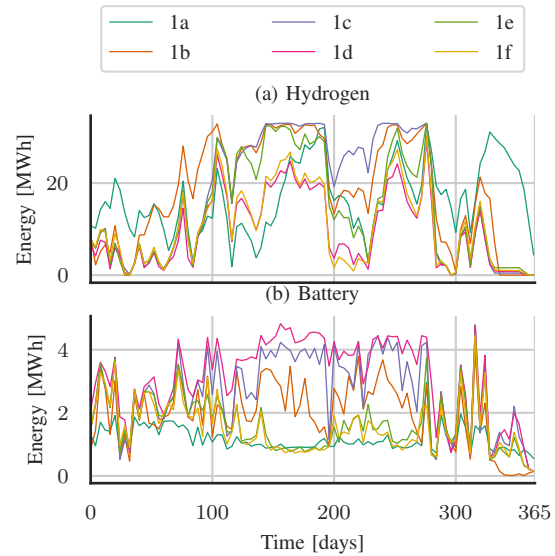}
    \caption{Four day average \gls{soc} for case \textit{1}.}
    \label{fig:avg_soc}
\end{figure}

\Cref{fig:avg_soc}a shows that the hydrogen \gls{soc} on average is lower when including the \gls{dod} degradation in methods \textit{d} and \textit{f}. The wind turbine peak capacity is 135 kW while the electrolyzer charge capacity is only 55 kW. In periods with very high wind power, the battery can act as a buffer for the hydrogen tank that is unable to absorb all the wind power. However, \gls{dod} degradation makes this less profitable, resulting in lower hydrogen filling.

\Cref{fig:avg_soc}b shows that the battery \gls{soc} for methods \textit{e} and \textit{f}, where \gls{soc} degradation is included, is stable low in mid-year when the demand is low. These periods require relatively low stored energy to secure the supply, especially when the solar \gls{pv} is delivering substantial energy due to many hours of sunlight through the summer. However, \gls{soc} degradation will, in general, lower the battery \gls{soc}, and \gls{dod} degradation will lower the hydrogen \gls{soc}, which in turn increases the risk of scarcity. The results in \cref{tab:case_cost_summary} shows a modest increase in both load shedding and diesel consumption when accounting for battery degradation.

\section{Conclusions} \label{sec:conclusions}
Battery degradation is strongly connected to the operational pattern. In this study, the expected battery lifetime was prolonged by more than four years by properly accounting for degradation effects caused by \gls{dod} and \gls{soc}. Moreover, the total operating costs were reduced by up to 25\% compared to the naive stochastic model without representation of degradation. However, battery degradation minimization also influences the operational pattern for the remaining resources in the system. The operational costs for generators and other \gls{ees} technologies can, in the worst case scenario, increase even more than the savings for battery degradation. Degradation costs and inefficiencies associated with the operational pattern should therefore be considered for the whole system. %

Dedicated battery degradation minimization can be contradictory to maximizing the security of supply, and the risk of scarcity must be balanced against the potential reduction in degradation. Realistic uncertainty models are therefore highly important.

The optimal operation of the future power system will to a greater extent be influenced by technology prices rather than fuel price, and energy adequacy rather than power adequacy. Future research should therefore give more attention to both the degradation of all the flexible resources in the system as well as precise uncertainty modeling to capture the future risk of scarcity accurately.

\bibliographystyle{IEEEtran}
\bibliography{references}

\end{document}